# Single Crystal High Entropy Perovskite Oxide Epitaxial Films

Yogesh Sharma[1], Brianna L. Musico[2], Xiang Gao[1], Chengyun Hua[3], Andrew F. May[1], Andreas Herklotz[1,4], Ankur Rastogi[1], David Mandrus[1,2], Jiaqiang Yan[1], Ho Nyung Lee[1], Matthew F. Chisholm[1], Veerle Keppens[2], T. Zac Ward[1*]

[1] *Materials Science and Technology Division, Oak Ridge National Laboratory, Oak Ridge, TN 37831, USA*
[2] *Department of Material Science & Engineering, University of Tennessee, Knoxville, TN 37996, USA*
[3] *Environmental and Transportation Science Division, Oak Ridge National Laboratory, Oak Ridge, TN 37831, USA*
[4] *Institute for Physics, Martin-Luther-University Halle-Wittenberg, Halle, Germany*
*wardtz@ornl.gov

**ABSTRACT**

The first examples of single crystal epitaxial thin films of a high entropy perovskite oxide are synthesized. Pulsed laser deposition is used to grow the configurationally disordered $ABO_3$ perovskite, $Ba(Zr_{0.2}Sn_{0.2}Ti_{0.2}Hf_{0.2}Nb_{0.2})O_3$, epitaxially on $SrTiO_3$ and MgO substrates. X-ray diffraction and scanning transmission electron microscopy demonstrate that the films are single phase with excellent crystallinity and atomically abrupt interfaces to the underlying substrates. Atomically-resolved electron energy loss spectroscopy mapping shows a uniform and random distribution of all B-site cations. The ability to stabilize perovskites with this level of configurational disorder offers new possibilities for designing materials from a much broader combinatorial cation pallet while providing a fresh avenue for fundamental studies in strongly correlated quantum materials where local disorder can play a critical role in determining macroscopic properties.

## I. INTRODUCTION

The $ABO_3$ perovskite oxide structure and its derivatives are of broad interest to the study and application of magnetism[1], energy conversion and storage[2,3], superconductivity[4], topology[5], ferroics[6–8], and a host of other phenomena. The perovskite's ability to produce such a varied range of functionalities lies in its structural and chemical flexibility, which can enable mixing cation combinations of vastly different character on the two different cation sublattices. Consequently, substitutional electron or hole doping on the A and B sites allows for a wide variety of charge and distortion states to be tuned through synthesis, with the cation size variance balanced by internal changes to Jahn-Teller distortions and octahedral tilts and rotations. This substitutional approach is a central pillar of materials design strategies—with the search for new functionally relevant materials often beginning with a parent $ABO_3$ ternary compound which is then partially substitutionally doped to an $[A_xA'_{1-x}]BO_3$ or $A[B_xB'_{1-x}]O_3$ quaternary compound of superior character or novel physical behavior [9,10]. Substitutional doping to quinary or higher states can provide further functional tunability or unexpected physics in strongly correlated systems, such as colossal magnetoresistance and emergent phase coexistence in $(La_xPr_yCa_{1-x-y})MnO_3$[11,12].

Intentional modification of long range structure through global isovalent substitution, transient pressure, heterostructuring, or strain effects[13–16] are widely used to induce changes to conduction pathways, spin states, and orbital degeneracies. Since many perovskite oxides have strongly correlated electrons, short and long range distortions can have different and dominating influences on behavior owing to the inherent spin-charge-lattice-orbital order parameter coupling length scales. The influence of local configurational disorder on mesoscopic properties is not well studied, but the existing examples are promising. As in the example of the quinary manganite above, modifying the type and magnitude of disorder can be a powerful tuning parameter in designing transition temperatures and magnetic phase compositions[17]. Further, manipulation of local structural disorder is a known route to restoring access to hidden quantum critical point phase spaces for the fundamental study of emergent behaviors[18,19]. The ability to create single crystal perovskites with very high levels of configurational disorder would open many new possibilities for materials design beyond simple electronic doping.

The lack of experimental studies on quinary or higher rank oxide perovskites is in large part due to the difficulty of stabilizing homogeneously doped single crystals. Typically, increasing the number of elements results in a higher probability of the formation of multiple phases or complicated microstructures. Approaches to predicting crystal stabilities are developing but are typically built upon calculations made at 0 K, which can be a severe limitation if entropy were to play a role in the stabilization process[20–22]. It was recently shown that entropy stabilized quinternary oxides, or high entropy oxides (HEOs), possessing a single cation sublattice could be synthesized[23]. The random distribution of constituent elements into the cation sublattice enhances the configurational entropy in such oxide solutions—analogous to the more well-known metallic high entropy alloys (HEAs)[24]. A range of rocksalt HEO compounds have been successfully synthesized in bulk single phase ceramic forms[23,25–31]. These results suggest that entropy stabilization can be especially effective in ionic compounds, hence the promise that it may be much easier to utilize entropic stabilization in oxides where the oxygen sublattice can be used to interrupt intermediate local electronegativity differences that can hinder stabilization of metallic HEAs[23,24]. Further, HEOs allow more complex crystal structures, such as spinels[31] and perovskites[32,33]. While ceramic forms of HEOs in the rocksalt, spinel, and perovskite phases have been stabilized, there is, thus far, only a single report of a HEO being stabilized in a single crystal form[34]. Here, the rocksalt structure was stabilized in epitaxial thin film form and, driven by the inherent local disorder of the HEO, shown to induce an order of magnitude increase in exchange coupling response at a ferromagnetic nickel-iron alloy interface. If such large disorder-mediated responses can be utilized in this relatively simple structure, the perovskite structure may offer even greater novelty of response due to its often extreme sensitivity to disorder.

In this work, we demonstrate the first example of a single crystal high entropy perovskite oxide (HEPO) by stabilizing the multicomponent $A(5B_{0.2})O_3$ perovskite $Ba(Zr_{0.2}Sn_{0.2}Ti_{0.2}Hf_{0.2}Nb_{0.2})O_3$ in epitaxial thin film form. This HEPO has a Goldschmidt tolerance factor of $t = 1.03$ which makes it an excellent candidate to stabilize in a cubic form[32]. The selection of B-site sublattice substitution is motivated by the general trend that changes to the oxidation state of the B-site and/or changes to the O-B-O and B-O bond lengths and angles often have a profound impact on perovskite functionality. Since the B-site sublattice is most often responsible for ferroic, magnetic, and electronic transport properties, the capability to select designer combinations of B-site stoichiometries offers new options to tailoring materials' properties and will likely lead to previously unobserved disorder-driven physical responses.

## II. EXPERIMENTAL DETAILS

A ceramic target of stoichiometric $Ba(Zr_{0.2}Sn_{0.2}Ti_{0.2}Hf_{0.2}Nb_{0.2})O_3$ was synthesized using the conventional solid-state reaction method[35]. HEPO thin films of varying thicknesses were then grown using pulsed laser deposition on 5×5×0.5 mm$^3$ $SrTiO_3$ (STO) and MgO single crystal substrates (CrysTec, Germany). These substrates were selected since there are no easily available substrates that lattice match the presented HEPO, thus we select substrates with smaller and larger lattice parameters. A KrF excimer laser (λ=248 nm) operating at 5 Hz was used for target ablation. The laser fluence was 0.8 J/cm$^2$ with an area of 3.5 mm$^2$ on the target. The target-substrate distance was set at 5 cm. Deposition optimization was performed, and the optimal growth conditions were found to occur with an oxygen partial pressure of 150 mTorr at a substrate temperature of 750 °C. After deposition, the films were cooled to room temperature under 100 Torr oxygen pressure. The growth rate per a laser shot was approximately 0.01 and 0.008 nm on STO and MgO substrates, respectively.

The crystal structure and growth orientation of the films were characterized by X-ray diffraction (XRD) using a four-circle high resolution X-ray diffractometer (X'Pert Pro, Panalytical) (Cu K$α_1$ radiation). Atomic force microscopy (Nanoscope III AFM) was used to monitor surface morphology of the as-grown films with all films showing <1 nm rms surface roughness. Cross-sectional specimens oriented along the [100] STO direction for scanning transmission electron microscopic (STEM) analysis were prepared using ion milling after mechanical thinning and precision polishing. High-angle annular dark-field (HAADF) and electron-energy loss spectroscopy (EELS) analysis were carried out in a Nion UltraSTEM 200 operated at 200 kV. An inner detector angle of about 78 mrad was used for HAADF observation, and a convergence angle of 30 mrad was used for EELS analysis. A Renishaw 1000 confocal Raman microscope was used to measure Raman spectra in back scattering configuration. The Raman spectrum was obtained from a sum average of 10 individual spectra taken at different places on the sample through a 20× objective. The wavelength of the Raman laser used in these measurements was 532 nm.

## III. RESULTS AND DISCUSSION

Three different thicknesses of Ba(Zr$_{0.2}$Sn$_{0.2}$Ti$_{0.2}$Hf$_{0.2}$Nb$_{0.2}$)O$_3$ films were grown on STO (001) substrates. Figures 1(a)-(c) show the x-ray diffraction results for each of the three films. $\theta$-$2\theta$ XRD scans demonstrate that all three thicknesses are single crystal c-axis oriented epitaxial films. There is no evidence of secondary phases even in the thickest film. Film uniformity is also excellent as demonstrated by Laue oscillations on the HEPO film 002 peak in Fig. 1(b) and the X-ray reflectivity (XRR) measurements where the periodic oscillations arising from the interfacial interference can be observed and fit to give thicknesses of 7, 26, and 72 nm. The full width at half maximum (FWHM) of the 002 rocking curve is 0.06° for the 26 nm film shown in Fig. 1(d), which, again, indicates an extremely high film quality. XRD $\phi$ scans on the 26 nm HEPO film show a cube-on-cube epitaxial relationship to the (001) oriented STO substrate with a characteristic 4-fold symmetry with a 90° spacing of the diffraction peaks, establishing the in-plane epitaxial relationship between the film and substrate as (001) HEPO || (001) STO ; [100] || [100] (Fig. 1(e)).

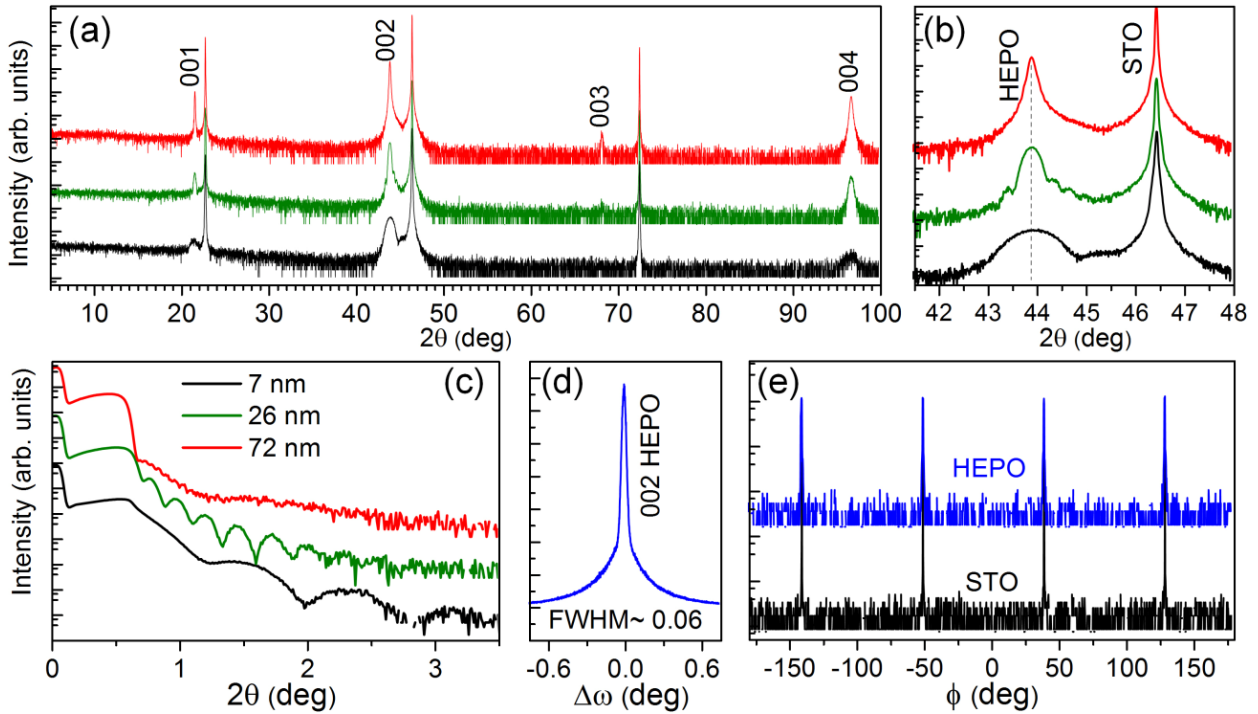

FIG. 1. X-ray diffraction results of the HEPO films on STO (001): (a) $\theta$-$2\theta$ XRD scans where the HEPO peaks are clearly labeled. (b) Enlarged view of XRD patterns around the 002 peaks of HEPO and STO. (c) XRR patterns of HEPO films with different thicknesses. (d) A rocking curve about the HEPO 002 peak ($\omega$ = 21.84) (e) $\phi$ scans of the 222 reflections of both film and substrate.

STO has in-plane lattice constants $a = b = 3.905$ Å which are significantly smaller than the expected cubic bulk lattice parameter found in the ceramic HEPO of $a = b = 4.115$Å[35]. This lattice mismatch of ~ 6% means that the films are unlikely to be coherently strained. Figure 1(b) shows that the all three film thicknesses align near the same value with the 7 nm film displaying some slight asymmetry in peak shape. To map films' coherency relationship to the underlying substrate, X-ray reciprocal space mapping (RSM) measurements were performed around the asymmetric (204) Bragg's reflection of the film and substrate (Fig. 2(a)-(c)). As expected, all films are relaxed from the substrate, as shown in the lack of vertical alignment of (204) film with respect to the substrate peak. Calculating lattice parameters from the x-ray data, we find the in-plane and out-of-plane lattice constants are nearly cubic for the thicker films with the 26 nm and 72 nm film's parameters $a = 4.122$ Å and $c = 4.121$ Å, which are near the expected bulk values. We find that some amount of compressive strain is passed to the 7 nm HEPO film. Its lattice parameters are measured as $a = 4.107$ Å and $c = 4.129$ Å. Assuming a cubic bulk value from those measured on the relaxed thick films $a_{bulk} \approx 4.1216$ Å, the 7 nm film is under crystal strains of $\varepsilon_{xx} = -0.354\%$ and $\varepsilon_{zz} = +0.179\%$. From these values, we can make a very rough initial estimate of the material's Poisson ratio ($v$) $\approx 0.2$, using $v = \varepsilon_{zz} / (\varepsilon_{zz} - 2\varepsilon_{xx})$.

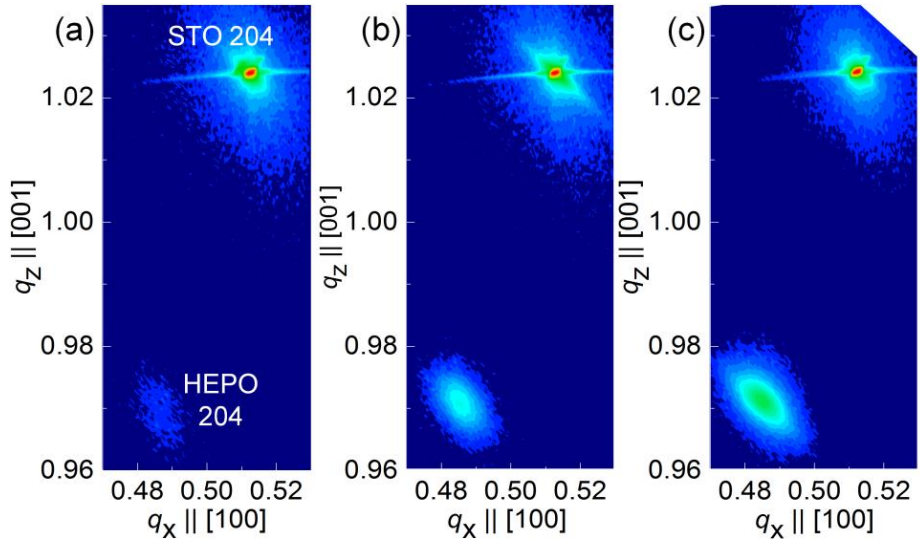

FIG. 2. Reciprocal space mapping (RSM) around the 204 reflections: (a) 7 nm; (b) 26 nm; and (c) 72 nm.

In order to examine the local distribution of the B-site constituent transition metal ions (Zr, Sn, Ti, Hf, and Nb) in the HEPO films, we conducted direct atomic level STEM imaging and spectroscopy. Figure 3(a) shows the high-angle annular dark field (HAADF) image of a single crystal 26 nm HEPO film on STO. Note that the light colored pockets on the film are the result of the sample preparation process where milled material collects on the surface in small amorphous blobs. The film shows a single crystalline lattice with an abrupt and fully coherent interface structure with the STO (001) substrate which is consistent with the RSM above where we see some strain passing to at least 7 nm (see also the magnified image in Fig. 3(b)). The STEM observations are also consistent with the XRD findings that the film is uniform and epitaxial. To confirm the local chemical homogeneity and distribution of B-site cations in the film, chemical analysis using atomically resolved EELS (STEM) was performed. Figure 3(c) shows a HAADF survey image where EELS measurements were performed. The EELS intensity maps of the Ti-L, Hf-M, Nb-M, Sn-M, and Zr-M signals are shown in Fig. 3(c). Consistent with the above measurements, the EELS mapping shows a uniform and random distribution of all cations throughout the observed region without any signs of cation segregation or clustering.

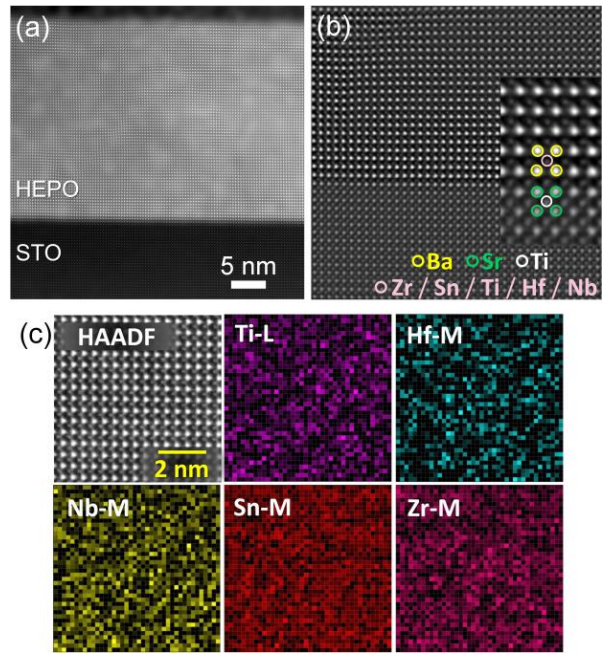

FIG. 3. STEM imaging and spectroscopy. (a) Cross-sectional HAADF-STEM observation of a HEPO/STO heterostructure along the [100] STO direction. (b) Magnified image of the film at the interface. A smooth interface with regular HEPO layers is observed. The inset shows a highest magnification with overlaid structural model and atomic columns. (c) HAADF survey image from the HEPO film grown on STO (001) where EELS mapping was performed with EELS elemental maps of Ti-L, Hf-M, Nb-M, Sn-M and Zr-M EELS signal from the region. The EELS mapping shows homogeneous distribution of B-site cations.

Ba(Zr$_{0.2}$Sn$_{0.2}$Ti$_{0.2}$Hf$_{0.2}$Nb$_{0.2}$)O$_3$ films were also grown on MgO (001) substrates. In comparison to STO (a = 3.905 Å), the MgO substrate (a = 4.212 Å) has a larger lattice parameter, which should permit the application of tensile strain to the film. Energy-dispersive spectroscopy (EDS) in scanning electron microscope (SEM) showed no evidence of quenched disorder and/or larger scale cationic clustering[35]. Figure 4(a)-(c) shows the XRD, XRR, and a RSM for three different thicknesses of films grown on MgO. Similar to those grown on STO, the films are all single phase and epitaxial to the substrate surface. RSM data was taken on the 20 nm and 62 nm films. The thinner film had lattice parameters of $a$ = 4.143 and $c$ = 4.113 and the thicker film's parameters were $a$ = 4.140 and c=4.116. Using the same bulk value ascertained from the relaxed STO thick film, we calculate the Poisson ratio to be ~0.16 and ~0.12 respectively. Thus we can estimate the Poisson ratio of Ba(Zr$_{0.2}$Sn$_{0.2}$Ti$_{0.2}$Hf$_{0.2}$Nb$_{0.2}$)O$_3$ as being within the range $\nu \approx 0.12 - 0.2$, which is below the range of 0.3 – 0.5 observed in the majority of ABO$_3$ perovskite oxides[36]. To better understand how the B-site's large configurational disorder impacts elastic and dynamic properties, we apply Raman spectroscopy and time-domain thermoreflectance (TDTR) to glimpse how symmetry and disorder may be impacting lattice vibration modes and thermal transport characteristics.

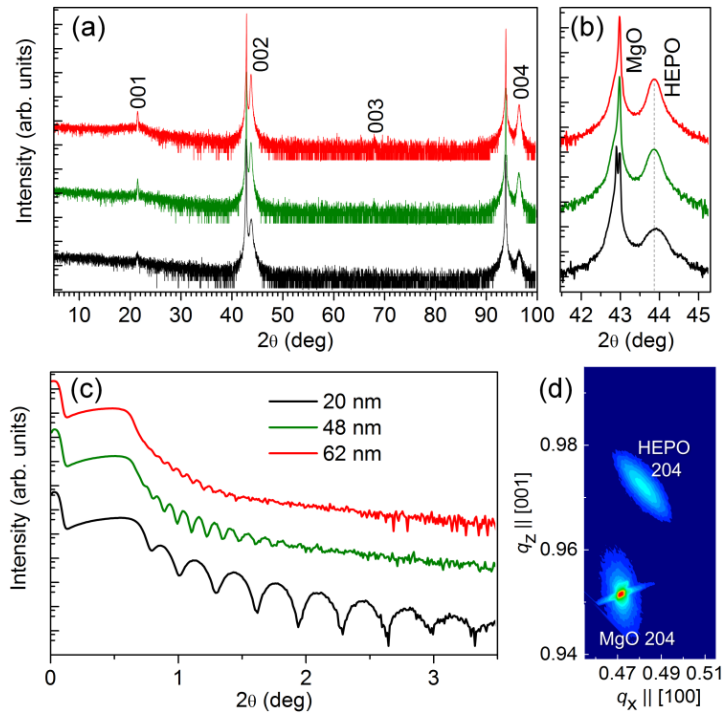

FIG. 4. X-ray diffraction of HEPO films on MgO (001). (a) θ-2θ XRD scans where the HEPO peaks are labeled. (b) Enlarged view of XRD patterns around the 002 peaks of HEPO and MgO.

(c) XRR patterns of HEPO films with different thicknesses. (d) Reciprocal space mapping (RSM) of 62 nm thick HEPO films taken around the 204 reflection.

We employ Raman spectroscopy to gain insight into the B-site cation symmetries and to gain a deeper look at cation disorder in our HEPO samples. We select the 62 nm film grown on MgO (001), as the MgO substrate provides a flat background without any strong Raman modes in the measured frequency range[37]. Figure 5(a) shows the room temperature micro-Raman spectrum of HEPO/MgO film. The Raman spectrum of our HEPO film resembles the spectra of thin films and bulk nanostructures of $BaTiO_3$ (BTO) and $Ba(Ti,Zr)O_3$ (BZTO)[38–40]. Based on the symmetry assignment, the Raman spectra confirm the presence of six phonon modes; $A_1(TO_1)$ ~171 cm$^{-1}$, $A_1(TO_2)$~253 cm$^{-1}$, $E(TO)$~312 cm$^{-1}$, $A_1(LO_2)/E(LO)$~437 cm$^{-1}$, $A_1(TO_3)$~512 cm$^{-1}$, $A_1(LO_3)/E(LO)$~ 730 cm$^{-1}$, corresponding to the tetragonal symmetry[38,39]. The high intensity mode at ~802 cm$^{-1}$, observed near the $A_1(LO_3)$ mode, is generally considered an indicator of relaxor behavior in doped-BTO or BZTO systems, due to the off-centering of B-site cations[39,41] while the broadening of the $A_1(TO_2)$ and $E(TO)$ modes likely indicates the presence of weak mode-coupling[42]. These observations are reasonable consequences of a system possessing random-site occupation of five cations of different ionic sizes. Functionally, the impact of such a large amount of disorder should subsequently have some impact on the ability of the lattice to transport heat due to modification to phonon modes.

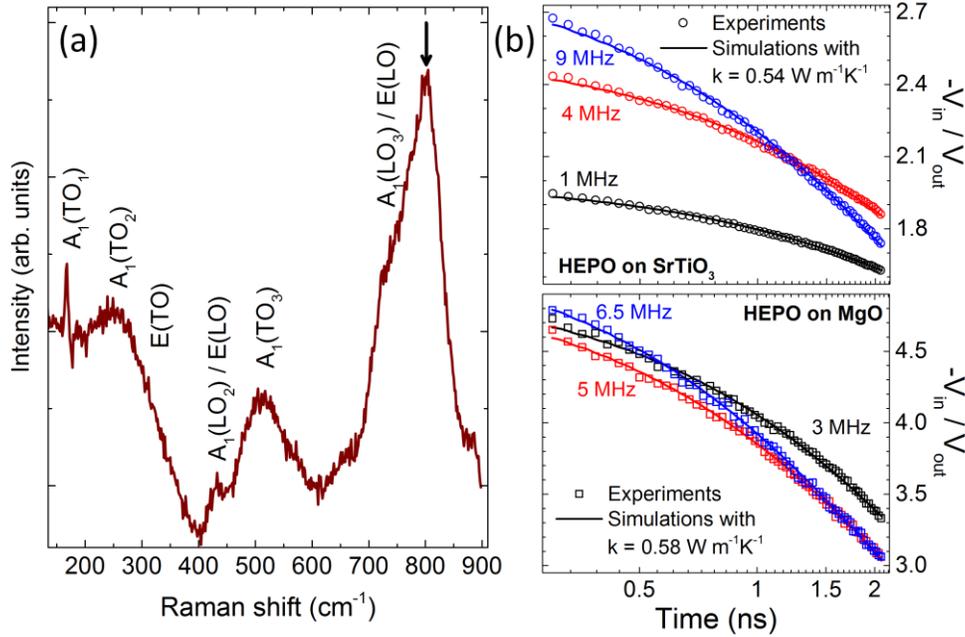

FIG. 5. Lattice vibration and thermal transport characteristics: (a)Room temperature unpolarized micro-Raman spectrum of 62 nm thick HEPO film grown on MgO(001). (b) Comparison of TDTR of HEPO films on STO and MgO substrates. The ratio of the in-plane to out-of-plane signals at three different modulation frequencies for each sample is plotted as a function of the delay time (t) between pump and probe. The symbols are experimental measurements and the lines are the corresponding theoretical fitting results for a set thermal conductivity (k) value as the one fitting parameter.

The thermal transport properties were studied using time-domain thermoreflectance (TDTR). Utilizing a two-tint pump-probe setup[43–45], we measured the thermal conductivity of 25 nm thick HEPO films grown on SrTiO$_3$ and MgO substrates. For TDTR transduction, the HEPO samples were coated with a 70 nm thick aluminum cap layer via e-beam evaporation prior to TDTR testing. The thermal conductivities of the MgO and SrTiO$_3$ substrates were measured separately using TDTR and use literature values for Al film and substrates' heat capacities at room temperature. At least five measurements were taken on each sample at different locations with modulation frequency varying from 1 to 9 MHz. TDTR data for the two samples at three different modulation frequencies and theoretical fits for room temperature are shown in Fig. 5(b). The fitting is based on a three component model[46]: the Al film, the HEPO film, and the substrate; the interface thermal resistances are included in the thermal resistance of the HEPO layer. The thermal conductivity of the film grown on MgO is 0.58 ± 0.03 W/m-K and that on SrTiO$_3$ is 0.54 ± 0.04 W/m-K. These values are nearly an order of magnitude lower than other single crystal perovskite oxides having only one or two different elements on the B-site lattice[47]. In fact, these values are very near the theoretical amorphous limit of cubic phase BaTiO$_3$ which is estimated at ~0.48 W/m-K[48,49]. Single crystal Ba(Zr$_{0.2}$Sn$_{0.2}$Ti$_{0.2}$Hf$_{0.2}$Nb$_{0.2}$)O$_3$ can thus be considered as having an ultralow, or glass-like, thermal conductivity. This is an interesting observation since the material still possesses a configurationally ordered A-site sublattice. This suggests that the theoretical predictions that highly configurationally disordered single crystals could provide an avenue to circumvent the amorphous limit may be valid[50]. By introducing configurational disorder on the A-site sublattice, thermal conductivity might be decreased to values well below the amorphous limit by further driving the necessary combined changes to local strain field and sublattice site-to-site mass differences which drive phonon scattering and limit heat flow.

**CONCLUSIONS**

Laser molecular beam epitaxy is shown to be an effective route to obtaining high quality single crystal high-entropy perovskite oxide (HEPO) thin films. This is demonstrated by growing epitaxial thin films of Ba(Zr$_{0.2}$Sn$_{0.2}$Ti$_{0.2}$Hf$_{0.2}$Nb$_{0.2}$)O$_3$ on SrTiO$_3$ and MgO substrates. X-ray diffraction and STEM imaging reveal that the films are single phase with excellent crystallinity and atomically abrupt interfaces to the underlying substrates. Direct atomic level STEM-EELS and macroscopic SEM-EDS confirm a uniform distribution of B-site cations in the films. Raman spectroscopy reveals weak phonon mode coupling and hints at the possibility of relaxor-like behavior. Time-domain thermoreflectance measurements show that this material has a thermal conductivity which is an order lower than the configurationally ordered BaTiO$_3$ parent material and approaches BaTiO$_3$'s amorphous limit in a single crystal form even though it possesses a fully configurationally ordered A-site sublattice.

Since ABO$_3$-type perovskites show such a wide range of physical properties, further studies of HEPOs are likely to lead to new functionalities due to their distinct highly-tunable chemistries. The ability to use entropy stabilization to introduce extreme configurational disorder opens new possibilities for designing materials from a much broader combinatorial cation pallet and should be of particular interest to fundamental studies in strongly correlated quantum materials where local disorder can play a critical role in determining macroscopic properties. As a final comment, the tunability of cation sizes should also allow very fine tuning of lattice parameters which may lead to the development of a new means of creating tailored substrates for epitaxial film growth.

## ACKNOWLEDGEMENTS

This work was supported by the Department of Energy (DOE), Office of Science, Basic Energy Sciences, Materials Sciences and Engineering Division. TDTR measurements were supported by the Laboratory Directed Research and Development Program of Oak Ridge National Laboratory, managed by UT-Battelle, LLC, for the U.S. Department of Energy. BLM acknowledges the support of the Center for Materials Processing, a Tennessee Higher Education Commission (THEC) supported Accomplished Center of Excellence. DM acknowledges support from the Gordon and Betty Moore Foundation's EPiQS Initiative through Grant GBMF4416. Research was in part conducted through user proposal at the Center for Nanophase Materials Sciences, which is a US DOE, Office of Science User Facility. Powder XRD was performed at the Joint Institute for Advanced Materials (JIAM) Diffraction Facility, located at the University of Tennessee, Knoxville.

This manuscript has been authored by UT-Battelle, LLC under Contract No. DE-AC05-00OR22725 with the U.S. Department of Energy. The United States Government retains and the publisher, by accepting the article for publication, acknowledges that the United States Government retains a non-exclusive, paid-up, irrevocable, world-wide license to publish or reproduce the published form of this manuscript, or allow others to do so, for United States Government purposes. The Department of Energy will provide public access to these results of federally sponsored research in accordance with the DOE Public Access Plan (http://energy.gov/downloads/doe-public-access-plan).